\documentclass[useAMS,usenatbib]{mn2e}
\usepackage{rotating}

 \usepackage{times}

\def\pcm3{{\rm\thinspace cm^{-3}}}

\def\contcaption{\@conttrue\SFB@caption\@captype}

\def\n_h{{\rm n_{H}}}

\def\NH1{{$N_{\rm HI}~$}}

\def\ga{{\rm\thinspace gauss}}

\def\approxlt{\mathrel{\hbox{\rlap{\lower .5ex \hbox {$\sim$}}
        \raise .15 ex \hbox{$<$}}}}
\def\approxgt{\mathrel{\hbox{\rlap{\lower .5ex \hbox {$\sim$}}
        \raise .15 ex \hbox{$>$}}}}

\def\la{\mathrel{\hbox{\rlap{\hbox{\lower4pt\hbox{$\sim$}}}\hbox{$<$}}}}
\def\ga{\mathrel{\hbox{\rlap{\hbox{\lower4pt\hbox{$\sim$}}}\hbox{$>$}}}}

\newbox\grsign \setbox\grsign=\hbox{$>$} \newdimen\grdimen
\grdimen=\ht\grsign
\newbox\simlessbox \newbox\simgreatbox \newbox\simpropbox
\setbox\simgreatbox=\hbox{\raise.5ex\hbox{$>$}\llap
     {\lower.5ex\hbox{$\sim$}}}\ht1=\grdimen\dp1=0pt
\setbox\simlessbox=\hbox{\raise.5ex\hbox{$<$}\llap
     {\lower.5ex\hbox{$\sim$}}}\ht2=\grdimen\dp2=0pt
\setbox\simpropbox=\hbox{\raise.5ex\hbox{$\propto$}\llap
     {\lower.5ex\hbox{$\sim$}}}\ht2=\grdimen\dp2=0pt
\def\simgreat{\mathrel{\copy\simgreatbox}}
\def\simless{\mathrel{\copy\simlessbox}}

\title[]{Praesepe and the seven white dwarfs}

\author[P. D. Dobbie et al.]{P. D. Dobbie$^{1}$\thanks{E-mail:
pdd@star.le.ac.uk}  D. J. Pinfield$^{2}$ R. Napiwotzki$^{1}$ N. C. Hambly$^{3}$ M. R. Burleigh$^{1}$ \newauthor M. A. Barstow$^{1}$ R. F. Jameson$^{1}$ and I. Hubeny$^{4}$\\
$^{1}$Department of Physics and Astronomy, University of Leicester, University Road, Leicester LE1 7RH, UK\\
$^{2}$Science \& Technology Research Institute, University of Hertfordshire,
College Lane, Hatfield, AL10 9AB \\
$^{3}$Wide Field Astronomy Unit, Institute for Astronomy, School of Physics, University of Edinburgh, Blackford Hill, Edinburgh, EH9 3HJ, UK\\
$^{4}$National Optical Astronomy Observatory, Tucson, AZ 85726, USA\\
}
\begin{document}

\date{Accepted 1988 December 15. Received 1988 December 14; in original form 1988 October 11}

\pagerange{\pageref{firstpage}--\pageref{lastpage}} \pubyear{2002}

\maketitle

\label{firstpage}

\begin{abstract}

We report the discovery, from our preliminary survey of the Praesepe open cluster, of two new 
spectroscopically confirmed white dwarf candidate members.  We derive the effective temperatures and surface
gravities of WD0837+218 and WD0837+185 (LB5959) to be $17845^{+555}_{-565}$K  and log g $= 8.48^{+0.07}_{-0.08}$
and $14170^{+1380}_{-1590}$K and log g $=8.46^{+0.15}_{-0.16}$ respectively. Using theoretical 
evolutionary tracks we estimate the masses and cooling ages of these white dwarfs to be 
$0.92\pm0.05$M$_{\odot}$ and $280^{+40}_{-30}$Myrs and $0.90\pm0.10$M$_{\odot}$ and $500^{+170}_{-100}$Myrs 
respectively. Adopting reasonable values for the cluster age we infer the progenitors of WD0837+218 and 
WD0837+185 had masses of $2.6\le$ M $\le $ M$_{\rm crit}$M$_{\odot}$ and $2.4\le$ M $\le 3.5$ M$_{\odot}$ respectively, where M$_{\rm crit}$ is the maximum mass of a white dwarf progenitor.
We briefly discuss these findings in the context of the observed deficit of white dwarfs in open clusters 
and the initial mass-final-mass relationship. 
\end{abstract}


\section{Introduction}

The common age, metallicity and distance of their members make galactic open star clusters 
favourable environments in which to examine fundamental issues in stellar and galactic astrophysics e.g. 
the shape of the initial mass function (IMF) or the form of the initial mass-final mass relationship 
(e.g. Weidemann 1987). The modestly rich and well studied  Praesepe (NGC2632) cluster at a distance of 177pc,
 as determined from Hipparcos astrometric measurements (Mermilliod et al. 1997), appears particularly 
suited to such investigations. Its members share a distinct proper motion so it is comparatively 
straightforward to discriminate them from the vast majority of field objects along this line of sight. 
For example, Hambly et al. (1995) performed an 
astrometric survey of 19 sq. degrees centered on the cluster and found the proper motions of members
tightly clumped around $\mu_{\alpha} cos~\delta = -30$ mas yr$^{-1}$ and $\mu_{\delta} = -8 $mas yr$^{-1}$.
A more recent Hipparcos based study of Praesepe finds mean values of $\mu_{\alpha} cos~\delta = 
-35.7$ mas yr$^{-1}$ and $\mu_{\delta} = -12.7 $mas yr$^{-1}$ (van Leeuwen 1999).

\begin{table*}
\begin{minipage}{175mm}
\begin{center}
\caption{Details of the six new candidate WD members unearthed (top) and the four previously known WD members recovered (bottom) by our preliminary survey of Praesepe.}
\label{mass1}
\begin{tabular}{llcccccccr}
\hline
Designation & Other ID  & RA & DEC & O & E & B$_{\rm J}$ & R$_{\rm F}$ &$\mu_{\alpha} cos~\delta$ & $\mu_{\delta}$ \\
& & \multicolumn{2}{|c|}{J2000.0} & \multicolumn{2}{|c|}{USNO-B} &\multicolumn{2}{|c|}{SuperCOSMOS} &\multicolumn{2}{|c|}{mas yr$^{-1}$} \\
\hline
candidate 1 &  -         &  08 36 10.01 & 19 38 19.1 & 17.60 & 18.17  & 18.44& 18.30& -36$\pm2$  & -2$\pm3$  \\  
WD0837+185 & LB5959     &  08 40 13.30 & 18 43 26.4 & 17.81 & 18.20  &18.28 & 18.24&  -32$\pm0$  & -12$\pm7$ \\
WD0837+218 &  -         &  08 40 31.47 & 21 40 43.1 & 17.87 & 18.64  &18.04 &18.17 &   -30$\pm5$ & -6$\pm5$  \\
candidate 4 &  -         &  08 42 58.03 & 18 54 35.5 & 17.86 & 18.44  & 18.34& 18.43&   -34$\pm3$ & -8$\pm2$  \\
candidate 5 &  -         &  08 43 22.00 & 20 43 30.5 & 18.34 & 18.39  & 18.44& 18.54&   -30$\pm4$ & -8$\pm2$  \\
candidate 6 & LB8648     &  08 46 01.91 & 18 30 48.5 & 18.04 & 18.25  & 18.11& 18.25&   -36$\pm5$ & -14$\pm5$ \\
\hline
WD0836+201 & LB393, EG61      &  08 39 45.57 & 20 00 16.0 & 17.54 & 18.24  & 17.98 & 18.15 &   -32$\pm2$ & -10$\pm4$ \\
WD0836+199 & LB1847, EG60     &  08 39 47.20 & 19 46 12.1 & 17.64 & 18.31  &18.73 &18.25 &   -38$\pm3$ & -10$\pm2$ \\
WD0837+199 & LB390, EG59      &  08 40 28.09 & 19 43 34.8 & 17.28 & 17.78  &17.48 &17.59 &   -34$\pm2$ & -2$\pm3$ \\
WD0840+200 & LB1876     &  08 42 52.32 & 19 51 11.3 & 17.27 & 18.00  &17.75 &18.02 &   -30$\pm3$ & -10$\pm4$  \\ 
\hline
\end{tabular}
\end{center}
\end{minipage}
\end{table*}
 
A spectroscopic study of F type members indicates the cluster is slightly metal 
rich with respect to the Sun ([Fe/H]=0.038, [C/H]=0.01; Freil \& Boesgaard 1992). This is 
consistent with the conclusions reached by previous investigations of this type (e.g. Boesgaard \& Budge 1988).
However, there is still uncertainty as to the age of the cluster, with estimates ranging from 0.4-2 
Gyrs (e.g. Allen 1973; Mathieu \& Mazeh 1988). Those 
determinations based on isochrone fitting generally support an age of between 0.7-1.1 Gyrs 
(e.g. Anthony-Twarog 1982; Mazzei \& Pigatto 1988), although Claver et al. (2001), hereafter C01, favour 
a value closer to that of the Hyades (625Myrs), on grounds that the two clusters have similar metalicity and, 
kinematically, Praesepe is part of the Hyades moving group (Eggen 1960).

To date, five white dwarf members of Praesepe have been identified: LB390, LB5893, LB393, LB1847 and LB1876 
(Luyten 1962; Eggen \& Greenstein 1965; Anthony-Twarog 1982, 1984; C01). This is fewer than
the 7-20 observable degenerates predicted from the extrapolation of the present day cluster luminosity function, 
allowing for reasonable assumptions about the form of the IMF, the maximum 
progenitor mass (M$_{\rm crit}$) and the binary fraction (Williams 2004, hereafter W04).
Several explanations have been put forward to account for this shortfall and the deficit of white dwarfs
observed in other open clusters such as the Hyades. For example, if  M$_{\rm crit}\sim4$M$_{\odot}$ there would have 
been fewer white dwarf progenitors in the first place (e.g. Tinsley 1974). However, the presence of the 
white dwarf LB1497 in the Pleiades with an estimated progenitor mass M$\simgreat6$M$_{\odot}$ 
argues against this (C01). It has recently been shown that asymmetry at a level of only 1\% in the post main 
sequence mass loss process is sufficient to lead to the rapid loss of a significant fraction of the white 
dwarf population from an open cluster (Fellhauer et al. 2003). Alternatively, for Praesepe at least, it 
may be that no investigation to date has included a sufficient fraction of the total area the 
cluster projects on the sky. The surveys of Anthony-Twarog (1982, 1984) and C01 have both 
concentrated on the central $\sim2$ sq. degrees of the cluster but Adams et al. (2002) determine 
the tidal radius to be $\sim5^{\circ}$ (see Figure 1).    

There have been a number of studies of the five previously known Praesepe white dwarfs. Reid (1996) have used high 
resolution spectroscopy of the H$-\alpha$ line cores to derive gravitational redshift based 
mass estimates of 0.42, 0.91 and 0.67M$_{\odot}$ for LB390, LB5893 and LB393 respectively. 
C01 fit synthetic line profiles to moderate resolution, high S/N spectra of the
H$-\eta$ to H$-\beta$ members of the Balmer series in each white dwarf, to measure effective temperatures 
(T$_{\rm eff}$) and surface gravities (log g). Subsequently, they derived the mass of each, 
in the order listed in the above
paragraph, to be 0.82, 0.91, 0.62, 0.82 and 0.75M$_{\odot}$. They also noted the anomolously low mass
 Reid determined for EG59 likely stemmed from his neglect of the Zeeman splitting of the H$-\alpha$ line 
core caused by a magnetic field with a strength of $\sim3$MG. Reid and C01 have estimated the mass of 
the progenitor star of each degenerate, comparing the difference 
between the age of the cluster and the cooling time for the white dwarf to the predictions of stellar 
evolutionary models. For four of the white dwarfs studied the progenitor mass is consistent with the 
existence of a monotonically increasing relationship between the initial mass and the final mass 
(see Figure 11 of C01). However, in both investigations LB5893 is found to be ``too young'' for its 
comparatively high mass. C01 speculate that it may be the outcome of binary evolution, perhaps a double 
degenerate merger. Alternatively, taking into account both this white dwarf and his mass estimate for LB390,
 Reid suggests that a simple relationship between initial mass and final
 mass may not exist. 

To move towards a resolution of these issues we are embarking on a comprehensive search for additional white 
dwarf members of Praesepe. Here we report the discovery, from a preliminary version of this 
survey, of two new white dwarf candidate cluster members. For each object we present an optical 
spectrum, determine T$_{\rm eff}$ and log g and by comparing these measurements to 
evolutionary models estimate mass and cooling time. We conclude by briefly discussing our findings in 
the context of the reported deficit of white dwarfs in Praesepe and the initial mass-final mass relationship. 
\vspace{0.1cm}
\section[]{A preliminary search for Praesepe white dwarfs}

We have utilised the USNO-B1.0 catalogue to undertake a survey of a $5^{\circ} \times 5^{\circ}$ region 
centred on the Praesepe open cluster ($\alpha=08~40$ $\delta= +19~40$, J2000.0). The USNO-B catalogue contains 
astrometric information and photographic magnitudes for over a billion objects, gleaned from digitally 
scanned photographic plates spanning a baseline of $\sim50$ years. The internal astrometric accuracy and the 
dispersion in the photometry are estimated to be $\sim0.2''$ and $\sim0.3$ magnitudes respectively (for details 
see Monet et al. 2003).

In this preliminary effort we have extracted all sources with $19 \ge $ O $\ge17$, O-E $ \le 0$ and with proper
motions $-25\ge \mu_{\alpha} cos~\delta \ge -45 $ mas yr$^{-1}$, $0 \ge \mu_{\delta} \ge -20$ 
mas yr$^{-1}$. This encompasses the magnitude range of known cluster white dwarfs and is virtually coincident
with the astrometric range sampled by Hambly et al. (1995). Further, the survey should
be near complete for O-E $ \simgreat -1$ (Hambly et al. 1995). Subsequently, the POSS II J and F images of each 
candidate have been inspected to eliminate extended sources, blended objects and spurious detections originating
in the diffraction spikes of bright stars. As an additional check, candidates have been cross referenced against the
2MASS Point Source Catalogue (Skrutskie et al. 1997), keeping only those which are either non-detections or have blue near-IR colours within the photometric errors. Finally, we have used photographic photometry measured by SuperCOSMOS
to compare the location of our new candidates to the locus of cluster white dwarfs in the B$_{\rm J}$, 
R$_{\rm F}$ colour-magnitude diagram (Figure 2). The external accuracy of individual passband magnitudes in the
SuperCOSMOS Sky Survey is quoted as 0.3 magnitudes, but this uncertainty is dominated by drifts in zeropoints as a 
function of magnitude and position on the sky. These systematic errors do not appear when using colours
as they are the same in all passbands, so indices like B$_{\rm J} - $ R$_{\rm F}$ are accurate to 
$\sim0.1$ magnitudes (Hambly et al. 2001). This is not the case in USNO-B where the scatter in 
colours is $>0.3$ magnitudes.  

To enhance the white dwarf sequence which is rather loosely defined by the five known degenerate 
members (the photometry for LB5893 appears to have been adversely affected by the proximity of the bright stellar 
cluster member KW195) we have used suitable objects drawn from the 20pc sample of Holberg et al. (2002) with 
trigonometric parallax determinations and SuperCOSMOS B$_{\rm J}$ and R$_{\rm F}$ photometry, scaling these 
to the cluster distance of 177pc. Of the seven new candidates the locations of six are deemed consistent 
with them being white dwarf members of Praesepe, the remaining object lying $\sim1$ magnitude below the sequence (see Figure 2).
Details of these six candidates and the four previously known white dwarf members recovered here are given in 
the top and the bottom of Table 1 respectively. Although C01 recognise LB5893 to be an astrometric member, it is
 not recovered here as its proper motion is listed in the USNO-B1.0 catalogue as $\mu_{\alpha} cos~\delta=-56$ 
mas yr$^{-1}$, $\mu_{\delta} =-14$ mas yr$^{-1}$. This astrometry also appears to have been affected by the
proximity of KW 195.

\begin{figure}
\vspace{200pt}
\includegraphics{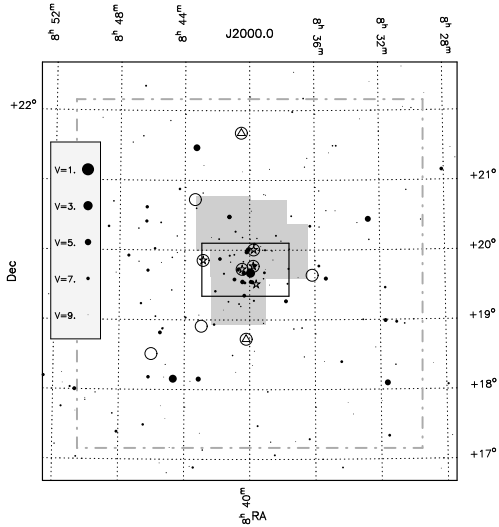}
\caption{A schematic plot of the Praesepe cluster showing stars down to V $\approx9$ and the 
areas surveyed by Anthony-Twarog (1982, 1984; solid outline) and C01 (grey shading). The region 
included in this investigation is outlined (dashed grey line). All objects listed in Table 1 
(open cricles) and the known white dwarf cluster members (open stars) are also overplotted.
The locations of the two new spectroscopically confirmed white dwarf candidate members are 
highlighted (open triangles).}
\end{figure}

\begin{figure}
\vspace{180pt}
\includegraphics{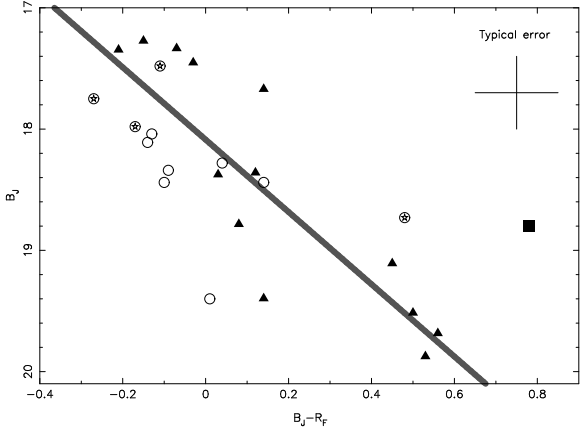}
\caption{A B$_{\rm J}$, B$_{\rm J} - $ R$_{\rm F}$ colour-magnitude diagram of the eleven objects remaining after cross referencing against the 2MASS PSC (open circles). The thick line represents a linear least squares fit to SuperCOSMOS photometry of the known degenerate members (open circles+stars) and objects drawn from the 20pc sample of Holberg et al. (2002) with trigonometric parallax determinations (filled triangles). The magnitudes of the latter have been scaled to correspond to a distance of 177pc. Note that LB5893 (filled square) has been excluded from the fit.}
\end{figure}

We have obtained optical spectra of two candidates in the sample (LB5959 and WD0837+218) using the William Herschel Telescope and the double 
armed ISIS spectrograph on 28/01/2001. Sky conditions were fair on this night with clear skies but with seeing 
$\sim2$''. For the course of the run ISIS was configured with the 5700 dichroic and the EEV12 and TEK4 detectors 
on the blue and red arms respectively. The data were obtained using the R158B and R158R gratings and a slit 
width of 1'' to provide a spectral resolution of $\approx6$\AA. Total exposure times were 60 and 120 minutes 
for LB5959 and WD0837+218 respectively. The CCD frames were bias subtracted,
flat fielded and cosmic ray hits removed using the IRAF routines CCDPROC and FIXPIX. Subsequently the spectra
were extracted using the APEXTRACT package and wavelength calibrated by comparison with the CuAr+CuNe arc spectra.
Our observed spectral standards (G191-B2B and Feige 67) were drawn from the catalogues of Oke (1974,1990) and used to
remove the instrument signature and telluric features from the science spectra. 

\begin{figure}
\vspace{128pt}
\includegraphics{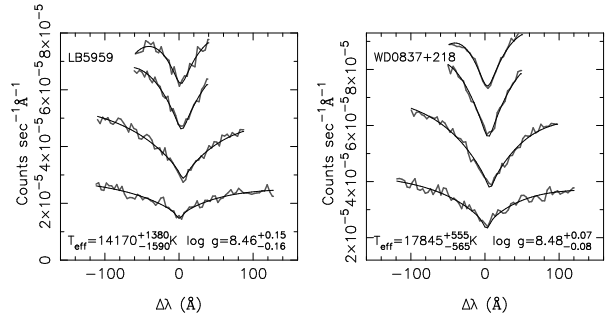}
\caption{The results of our fitting of synthetic profiles (thin black lines) to the observed Balmer lines (thick grey lines).}
\end{figure}

\section{Analysis of the data} 

\subsection{Model white dwarf spectra}

A glance at Figure 3 reveals broad hydrogen Balmer lines consistent with both objects being DA 
white dwarfs. 
Therefore, we have generated a grid of pure-H synthetic spectra covering the T$_{\rm eff}$ and surface 
gravity ranges 14000-20000K and log g=7.0-9.0 respectively. We have used the latest versions of the plane-parallel, 
hydrostatic, non-local thermodynamic equilibrium (non-LTE) atmosphere and spectral synthesis codes TLUSTY (v200; Hubeny
1988, Hubeny \& Lanz 1995) and SYNSPEC (v48; Hubeny, I. and Lanz, T. 2001, ftp:/tlusty.gsfc.nasa.gov/synsplib/synspec).
We have employed a state-of-the-art model H atom incorporating the 8 lowest energy levels and one superlevel extending 
from n=9 to n=80, where the dissolution of the high lying levels was treated by means of the occupation probability 
formalism of Hummer \& Mihalas (1988), generalised to the non-LTE situation by Hubeny, Hummer \& Lanz (1994). All 
calculations were carried out under the assumption of radiative equilibrium, included the bound-free and free-free 
opacities of the H$^{-}$ ion and incorporated a full treatment for the blanketing effects of HI lines and the 
Lyman $-\alpha$, $-\beta$ and $-\gamma$ satellite opacities as computed by N. Allard (e.g. Allard et al. 2004).
During the calculation of the model structure the lines of the Lyman and Balmer series were treated by means of an 
approximate Stark profile but in the spectral synthesis step detailed profiles for the Balmer 
lines were calculated from the Stark broadening tables of Lemke (1997). 

\begin{table}
\begin{minipage}{85mm}
\begin{center}
\caption{Details of the two new spectroscopically confirmed white dwarf candidate cluster members.
Masses and cooling times are derived from the ``thick H-layer'' evolutionary calculations of Wood (1995). }
\label{wdmass}
\begin{tabular}{lcccc}
\hline
WD   & T$_{\rm eff}$(K) & log g & M(M$_{\odot}$) &  $\tau_{c}$(Myrs)\\
 \hline

0837+185          & $14170^{+1380}_{-1590^{*}}$ & $8.46^{+0.15}_{-0.16}$ &  $0.90\pm0.10$ & $500^{+170}_{-100}$ \\
\\
0837+218          & $17845^{+555}_{-565}$ & $ 8.48^{+0.07}_{-0.08}$ &  $0.92\pm0.05$ & $280^{+40}_{-30}$ \\
\hline
\end{tabular}
\end{center}
 $^{*}$ extrapolated.
\end{minipage}
\end{table}

\subsection{Determination of effective temperatures and surface gravities}

We carried out comparisons between models and data using the spectral fitting program XSPEC (Shafer et al. 1991). XSPEC works by folding a model through the instrument response before comparing the result to the data by means of a $\chi^{2}-$statistic. The best fit model representation of the data is found by incrementing free grid parameters in small steps, linearly interpolating between points in the grid, until the value of $\chi^{2}$ is minimised. Errors are calculated by stepping the parameter in question away from its optimum value until the difference between the two values, $\Delta\chi^{2}$, corresponds to $1\sigma$ for a given number of free model parameters (e.g. Lampton et al. 1976). The errors in the T$_{\rm eff}$s and log g s quoted here are formal $1\sigma$ fit errors and may underestimate the true uncertainties. 

Preliminary fitting of our model grid to the observed Balmer line profiles (H$-\epsilon$ - H$-\alpha$) in both spectra revealed that our efforts to remove the effects of the wiggles in the response of the ISIS dichroic around 4400\AA\ had not been entirely successful. Therefore, we excluded the Balmer$-\gamma$ line from our subsequent analyses, determining T$_{\rm eff}$s and log g s from the four remaining profiles. The results are given in Table 2 and shown overplotted in Figure 3.

\section{Discussion}

\begin{table*}
\begin{minipage}{175mm}
\begin{center}
\caption{Progenitor lifetimes and corresponding masses for various adopted cluster ages. }
\label{pmass}
\begin{tabular}{lcccccc}
\hline
 Progenitor & \multicolumn{2}{|c|}{8.80 (log$_{10}$ Myrs)}    &  \multicolumn{2}{|c|}{8.92 (log$_{10}$ Myrs)}   & \multicolumn{2}{|c|}{9.04 (log$_{10}$Myrs)} \\
  of WD & \multicolumn{1}{|c|}{$\tau_{\rm prog}$ (Myrs)} & \multicolumn{1}{|c|}{M$_{\rm prog}$ (M$_{\odot}$)}  &  \multicolumn{1}{|c|}{$\tau_{\rm prog}$ (Myrs)} & \multicolumn{1}{|c|}{M$_{\rm prog}$ (M$_{\odot}$)} & \multicolumn{1}{|c|}{$\tau_{\rm prog}$ (Myrs)} & \multicolumn{1}{|c|}{M$_{\rm prog}$ (M$_{\odot}$)} \\
\hline
0837+185     & $130^{+100}_{-170}$ & $4.9^{\simless\rm{M}_{\rm crit}}_{-1.0}$ & $330^{+100}_{-170}$ & $3.4^{+1.1}_{-0.3}$  & $600^{+100}_{-170}$ & $2.8^{+0.3}_{-0.2}$  \\
0837+218  & $350^{+30}_{-40}$ & $3.3^{+0.2}_{-0.1}$ &  $550^{+30}_{-40}$ & $2.9^{+0.0}_{-0.1}$ & $820^{+30}_{-40}$ & $2.5^{+0.0}_{-0.1}$  \\
\hline
\end{tabular}
\end{center}
\end{minipage}
\end{table*}

\subsection{Cluster membership and the white dwarf deficit}

A steep powerlaw ($\Gamma=2$) was the only shape of IMF of the four investigated by W04 
consistent with his assumption of seven white dwarf members of Praesepe, for reasonable values
of the maximum progenitor mass (6M$_{\odot}\le$ M$_{\rm crit}\le10$M$_{\odot}$). The ``Naylor''
IMF, a broken powerlaw with index $\Gamma=0.2$ (M$\le1$M$_{\odot}$) and $\Gamma=1.8$ 
(M$>1$M$_{\odot}$) was found to be consistent only for M$_{\rm crit}=6$M$_{\odot}$. However, 
in a subsequent paper, Williams, Bolte \& Liebert (2004), show LB6037 and LB6072 to be QSOs, 
reducing the number of bona fide white dwarf members of Praesepe unearthed to date to only five.
In this case, only the steep powerlaw form with M$_{\rm crit}\le8$M$_{\odot}$ can be 
considered consistent with the observations.  

Nevertheless, we have presented evidence here the number of Praesepe white dwarfs is at least 
seven. As a further check of the membership status of our two new spectroscopic candidates we 
constrain their distances using the measured T$_{\rm eff}$s and log g s and radii derived from 
evolutionary tracks. From the ``thick H layer'' models of Wood (1995) we determine 
R$=0.0093\pm0.0020$R$_{\odot}$ and R$=0.0091\pm0.0009$R $_{\odot}$ for LB5959 and WD0837+218 respectively. 
Subsequently,we estimate the absolute visual magnitude of LB5959 to be M$_{\rm V} = 12.0\pm0.5$  and 
that of WD0837+218 to be M$_{\rm V} = 11.7\pm0.2$. Refering to the synthetic photometry of Bergeron et al.
 (1995; private comm.), we estimate B-V $\approx +0.15$ and +0.05 for the cooler and hotter degenerate 
respectively. Thus from the SuperCOSMOS data, B$_{\rm J}=18.3\pm0.3$ and B$_{\rm J}=18.0\pm0.3$, 
we determine V magnitudes of $18.15\pm0.3$ and $18.0\pm0.3$. Hence, we estimate these white dwarfs 
to reside at $170^{+45}_{-40}$pc and $180^{+40}_{-30}$pc respectively. 

Indeed, the favourable success rate of our preliminary survey in unearthing cluster white dwarfs 
suggests several of the remaining four new candidates will also prove to be degenerate members. 
However, as confirmation of this must await further spectroscopy, for now we consider the 
observed number of Praesepe white dwarfs, N, to lie in the range $7\le$ N $\le11$. Based on the 
Williams simulations we find the steep powerlaw form of the IMF is consistent with the 
observed number for any reasonable value of M$_{\rm crit}$. Similarly, if N$\ge8$   
the ``Naylor'' form can be considered consistent  for any reasonable 
value of M$_{\rm crit}$. We note, one requires to detect at least ten white dwarf members for the 
Salpeter form of the IMF to be regarded at best, in Williams scheme, as mildly inconsistent with 
the observed number (P$\approx0.07$). As our preliminary results indicate white dwarf members are 
to be found beyond the well studied inner regions of the cluster (Figure 1 shows WD0837+218 lies at
a projected separation of $\sim2^{\circ}$) more likely await discovery. A detailed survey extending
out to at least the tidal radius should be undertaken before any firm conclusions are drawn regarding 
the form of the IMF.

Prior to this work only the ``Naylor'' form of the IMF was found to be consistent with the non-detection 
of cluster white dwarfs residing in unresolved binary systems (W04). However, there is no evidence to suggest
that either of the two new confirmed white dwarfs resides in a binary. We find a white dwarf with 
T$_{\rm eff}=17000$K and log g $=8.0$, typical of the Praesepe population, has M$_{\rm V}$=11.0, 
M$_{\rm B}$=11.0 and  M$_{\rm U}$=10.2 (Bergeron et al. 1995) and a young disc M3 dwarf has M$_{\rm V}$=10.7, M$_{\rm B}$=12.3 and M$_{\rm U}$=13.4 (Leggett et al. 1992). An unresolved binary consisting of these two objects has 
U-B$=-0.6$, B-V$=0.6$. It thus lies on the fringes of the W04 criteria for being ``observable'' 
(U-B $\le0.0$, B-V $\le0.6$), which were chosen to approximate the limits of the UBV surveys of C01 and Anthony-Twarog. White dwarfs with more massive main sequence 
companions are unlikely to have been identified by these surveys. Similarly, since both Luyten's and the current 
work utilised blue and red photographic plates, selecting objects with colours I.C.$\le0.2$ ($\approx$ B-V$\simless0.5$) and O-E$\le0$ respectively, these too are likely biased against finding white dwarfs 
residing in binaries with stars of spectral type earlier than mid-M.
 
Farihi et al (2003) report a deficit of objects of spectral type later than mid-M paired to field white dwarfs at separations of $\sim$ few 100AU.
Further, radial velocity surveys point towards a drop in the relative frequency 
of main sequence binaries with mass ratios M$_{2}/$M$_{1}\simless0.2$, at separations $\simless5$ AU (Halbwachs et al. 2003; 
Marcy \& Butler 2000). Hence, it seems plausible, particularly as the progenitors of the Praesepe and Hyades 
white dwarfs had M$\simgreat2.5$M$_{\odot}$, that by including a population of zero age binaries consisting of 
randomly paired stars, the simulations of W04 overpredict the number of detectable white dwarfs in unresolved 
binaries with main sequence companions. This assumption was acknowledged by W04 as a possible shortcoming in 
their modelling. Our forthcoming GALEX survey 
of Praesepe will expand the parameter space in which we are able to search for white dwarfs to include those 
 in unresolved systems with K, G, and F type companions. It will reveal important additional information 
on the form of the IMF and the binary fraction of this cluster.

\subsection{White dwarf masses and the initial mass-final mass relationship}

We have estimated the masses of the two new white dwarfs by comparing their measured T$_{\rm eff}$s 
and log g s to the predictions of evolutionary calculations (Wood 1995). As expected for their 
progenitor masses, M $\simgreat2.5$M, these two objects have masses greater than the canonical white dwarf value of M $\approx 0.6$M$_{\odot}$ (see Table 2). We have also used the evolutionary tracks to estimate the cooling times of these 
objects, determining $500^{+170}_{-100}$Myrs and $280^{+40}_{-30}$Myrs for LB5959 and WD0837+218 respectively.
As the age of the cluster is rather uncertain we have used three different estimates encompassing the likely value
(log $\tau_{\rm cluster} = 8.80, 8.92 ~{\rm and}~ 9.04$) to derive the progenitor lifetimes.  Using cubic 
splines to interpolate between the lifetimes calculated for stars of solar composition by Girardi 
et al. (2000), we constrain the masses of the progenitors of LB5959 and WD0837+218 to be 
$2.6\le$ M $\le $ M$_{\rm crit}$ M$_{\odot}$ and $2.4\le$ M $\le 3.5$ M$_{\odot}$ respectively.

Examining the location of these objects in Figure 11 of C01 we find while LB5959 fits in comfortably
with a monotonic relationship between initial mass and final-mass, like LB5893, WD0837+218 appears to 
be too hot and hence too young for its high mass. We note there are five blue straggler members of Praesepe (e.g. Andrievsky 1998). 
The evolution of these objects appears to have been delayed, either through binary interaction or another as 
yet unidentified mechanism. We suggest that LB5893 and WD0837+218 may be related to this population. Alternatively, 
as suggested by Reid (1996), perhaps, at least some stars, do not subscribe to a simple monotonic 
relationship between their mass and the mass of their resulting white dwarf remnant.

\section*{Acknowledgments}
PDD and DPI are sponsored by PPARC postdoctoral grants. RN and MBU acknowledge the support of PPARC 
advanced fellowships. The WHT is operated on the island of La Palma by 
the Isaac Newton Group in the Spanish Observatorio del Roque de los Muchachos of the Instituto de
 Astrofisica de Canarias. This publication makes use of data products from the 
Two Micron All Sky Survey, which is a joint project of the University of Massachusetts and the Infrared 
Processing and Analysis Center/California Institute of Technology, funded by the National Aeronautics and
Space Administration and the National Science Foundation. We thank the anonymous referee for timely and useful 
comments.

\appendix

\bsp

\label{lastpage}

\end{document}